\documentclass{article}
\usepackage{arxiv}

\usepackage[utf8]{inputenc} 
\usepackage[T1]{fontenc}    
\usepackage{hyperref}       
\usepackage{url}            
\usepackage{booktabs}       
\usepackage{amsfonts}       
\usepackage{nicefrac}       
\usepackage{microtype}      
\usepackage{lipsum}
\usepackage{graphicx}
\usepackage{amssymb}
\usepackage{subfig}
\usepackage{natbib}
\graphicspath{ {./images/} }

 \title{Dynamical origin of Dimorphos from fast spinning Didymos}
 \author{Gustavo Madeira \\
 Universite de Paris, Institut de Physique du Globe de Paris, CNRS
 Paris, F-75005, France
\\
  Grupo de Dinamica Orbital de Planetologia, Sao Paulo State University (UNESP)
 Guaratingueta, SP 12516-410, Brazil\\
 \texttt{madeira@ipgp.fr}
 \And
 S\'ebastien Charnoz\\
 Universite de Paris, Institut de Physique du Globe de Paris, CNRS
 Paris, F-75005, France 
\And
Ryuki Hyodo\\
ISAS/JAXA, Sagamihara
Kanagawa, Japan
 }

\begin{document}
\maketitle
\begin{abstract}
Didymos is a binary near-Earth asteroid. It is the target of the DART and HERA space missions. The primary body, Didymos, rotates close to the spin at which it is expected to shed mass. The secondary body, Dimorphos, is a 140~meters moon that orbits the primary body in about 12~hours. Here we investigate the possible origin of Dimorphos. Using 1D models of ring/satellite interactions, we study the evolution of material lost from Didymos' surface and deposited as a ring at its equator. We find that due to viscous spreading, the ring spreads outside the Didymos' Roche limit forming moonlets. A fraction of the mass will form Dimorphos and a set of objects near the Roche limit, while most of the ring's mass falls back on Didymos. To match the properties of today's Dimorphos, the total mass that must be deposited in the ring is about $25\%$ of Didymos' mass. It is possible that a fraction of the material travelled several times between the ring and the surface of Didymos. The models produce an orbit similar to that observed for a Didymos tidal parameter $k_2/Q\leq10^{-5}$. If the ring deposition timescale is long ($\geq10^2$~yr) (so the material flux is small) Dimorphos could be irregularly shaped as it forms from the collision of similar-sized satellitesimals. However, the top-shape of Didymos is expected to be achieved due to a fast spin-up of the asteroid, which would result in a short deposition timescale ($\lesssim$yr). In that case, the satellite would form from progressively accreting material at the Roche Limit, resulting in an ellipsoidal Dimorphos constructed of small pieces with sizes of the order of meters, which is apparently in agreement with the recent images of Dimorphos obtained by DART mission.  
\end{abstract}

 keywords can be removed

\section{Introduction} \label{sec:intro}
Discovered in the 90s, (65803) Didymos is a near-Earth asteroid (NEA) and target of the NASA's DART mission. Designed to be an astronomical-scale kinetic impact test, the mission was launched in November 2021 \citep{cheng2018} and impacted Didymos' secondary, Dimorphos, in September 2022. The ESA mission HERA is scheduled to be launched in October 2024 and will also visit Didymos system. The goal of this mission is to do a post-impact characterization of the system \citep{michel2018}. Over the years, Didymos has been observed through radar and light-curve observations \citep{michel2016,naidu2016}, which allowed the characterization of shape and physical parameters of the object. Didymos is a S-type asteroid \citep{deLeon2010,Pravec2012} and has a top-shape appearance, resembling a spherical body with an equatorial ridge. \cite{naidu2020} evaluated the average object radius as $R_D=390$~m and spin period of $T_D=2.26$~h, while the bulk density is the least constrained physical parameter of the system ($\rho_D=2170\pm350~{\rm kg/m^3}$). The physical properties of Didymos and its secondary Dimorphos are summarised in Table~\ref{tab:physical}\footnote{We assume that Dimorphos is in a tidally locked state \citep{agruda2022}}.
\begin{table}{}
\caption{Physical properties of Didymos and Dimorphos \citep{Hirabayashi2017,naidu2020,Terik2022}\label{tab:physical}}
\centering
\begin{tabular}{llclc}
\hline\hline
Parameters & & Didymos  &  & Dimorphos \\ \hline
Principal semi-axes (km) & $\alpha_D$ & $0.40$ & $\alpha_d$ & $0.10$ \\
 & $\beta_D$ & $0.39$ & $\beta_d$ & $0.08$ \\
 & $\gamma_D$ & $0.38$ & $\gamma_d$ & $0.07$ \\
Mass (kg) & $M_D$ & $5.12\times 10^{11}$ & $M_d$ & $4.92\times 10^{9}$  \\
Bulk density (kg/m$^3$) & $\rho_D$ & $2170$ & $\rho_d$ & $2170$  \\
Average radius (km) & $R_D$ & $0.39$ & $R_d$ & $0.08$  \\
Spin period (h) & $T_D$ & $2.26$ & $T_d$ & $12.15$  \\
Orbital distance (m) &  & -- & $a_d$ & $1183$ \\
\hline
\end{tabular}
\end{table}

Didymos, such as 2001 SN263 \citep{becker2015}, 1999 KW4 \citep{Ostro2006}, and 1994 CC \citep{Brozovic2011}, is in a class of binary asteroids in which a top-shaped primary rotates near its spin limit \citep{Warner2009,margot2015,walsh2015}. When close to this limit, material on the surface of the object feels a centrifugal force that may overcome the gravitational attraction \citep{Pravec2007}, resulting in geological phenomena such as landslides \citep{walsh2008,Scheeres2015} or shape deformation \citep{Hirabayashi2014}. The Yarkovsky–O’Keefe–Radzievskii–Paddack (YORP) effect is a well-known effect responsible for accelerating the rotation of small bodies on a timescale of $\sim10^4-10^6$~yr, due to the re-radiation of the sunlight absorbed by the asteroid \citep{Bottke2006}. Given this, \cite{Scheeres2006} propose that when accelerated by the YORP effect, surface material flows towards the equator, which would result in the top shape observed in the asteroids rotating near the spin limit. Given this, one might ask whether material is put in orbit due to this process, giving rise to moons.

It is shown by \cite{walsh2008} that asteroids with high-friction angle may eject particles when rotating near the critical spin. Material dislodged in the ejection is believed to be deposited in low-eccentricity orbits around the object, giving rise to a satellite (secondary). However, the direct modelling of the friction forces was not realised and the effects of cohesion were not included by \citeauthor{walsh2008} in their hard-sphere approach. \cite{sugiura2021} goes a step further by adopting the effective bulk friction, which can effectively include the effect of cohesion. They demonstrate that, in fact, an initially spherical asteroid can achieve a top-shaped figure and form a transient ring in its vicinity. 

\cite{hyodo2022} performed longer-term simulations of \cite{sugiura2021}. They showed that a mass on the order of $10\%$ of the initial spherical body is ejected in a single avalanche event around the top-shaped central rubble-pile body, forming a transient particle disk followed by formation of rubble-pile moons via disk spreading. This may be the origin of Didymos' moon Dimorphos, the focus of our work.

\cite{Zhang2017,Zhang2018,Zhang2021} use the gravitational N-body code PKDgrav to study Didymos' spin and shape evolution \citep{Richardson2016}. Assuming Didymos as a rubble pile object, they found that a minimum cohesion of $10$~Pa is required for Didymos to maintain its structural stability and have no material detached, although its actual stability depends on the material arrangement and the density distribution within the body. The mass shedding phenomenon on Didymos was studied in \cite{Yu2018}. The increase in the asteroid's spin rate increases the mass detachment on Didymos and also affects the evolution of the lofted material. A slowly rotating Didymos is capable of trapping material for at least a few months in its vicinity, possibly promoting the formation of large debris via accumulation (possible Dimorphos' building blocks). As the spin rate increases, the stable region in the vicinity of the object becomes smaller, releasing trapped material, which is in agreement with the results of \cite{walsh2008} and \cite{hyodo2022}.

Another proposed mechanism for the formation of binary systems that can give rise to Dimorphos is the primary fission due to the rapid spin, which occurs for sufficiently cohesive bodies \citep{Pravec2010}. In such objects, the cohesion holds the object's shape until a break-up threshold energy is reached and a large fragment is detached from the primary, resulting in the satellite. If Didymos is cohesive enough, it may have fragmented into larger fragments in the past, meaning that Dimorphos possibly formed through this process. Now, if Didymos is a low cohesion object it is expected that material has been deposited at Didymos' equator due to avalanches on the object. In this case, Dimorphos can be formed by the accretion of the ejected material \citep{walsh2008,hyodo2022}.

In this work, we study the formation of Dimorphos from the material ejected from Didymos due fast spinning \citep{walsh2008,hyodo2022}. We study the viscous evolution of the ring of ejected material, with 1D numerical simulations using the \texttt{HYDRORINGS} code \citep{charnoz2010,salmon2010}. Here, our simulations are much longer timescale than previous studies, and we include the effect of tidal evolution of the formed moons. The effect of Didymos shape and the mass shedding process are not studied, which would require a more adequate tool. We assume that the mass deposition rate in the ring follows an e-folding temporal function. 

The paper outline is as follows. In Section~\ref{sec_dynamical}, we describe our dynamic model and numerical simulations. The orbital evolution of a system composed by Didymos and a ring of material is analysed in Section~\ref{sec:ring}. In Section~\ref{sec_parameters}, we analyse the effects of flow and tidal parameters and in Section~\ref{sec:grow}, we focus on the formation pathways of Dimorphos. In Section~\ref{sec_discussion}, we discuss the possible implications of our results on Dimorphos shape and the limitations of our model. We address our conclusions in Section~\ref{sec_conclusion}.

\section{Dynamic Model} \label{sec_dynamical}
Top-shaped structures are thought to be formed through deformation due to rapid rotation \citep{watanabe2019}, being the formation of ring the consequence of such a process \citep{walsh2008,Yu2018}. The spin period of Didymos today is near the spin limit, suggesting a possible formation of its top shape through a fast spin-up \citep{Zhang2017,Zhang2018,Zhang2021}, which could imply the formation of Dimorphos from a ring. Recently, \cite{sugiura2021} performed Smoothed Particle Hydrodynamics (SPH) simulations of granular bodies to model a rotational deformation of rubble-pile bodies. They showed that the top shape formation occurs through an axisymmetric set of surface landslides of a primary, induced by a fast spin-up. This landslide, then, results in a mass ejection around the primary, forming a debris ring followed by a moon formation \citep{hyodo2022}.

Based on this, we envision the following scenario for the formation of Dimorphos: the fast rotation of Didymos caused by an external effect, e.g. the YORP effect, induces avalanches on the object \citep{Yu2018}. Such an event is responsible for the asteroid reshaping and deposition of material in the equatorial plane (ring of material). Due to angular momentum conservation, Didymos spin rate decreases \citep{sugiura2021}. The ring spreads viscously and satellites are formed while part of the ring material falls on the primary \citep{charnoz2011,hyodo2015,madeira2022}, which increases its rotation. 

\cite{Zhang2021} and \cite{sugiura2021} find that Didymos reshaping would occur via a single avalanche, with material that falls back to the body being reassembled by it. However, due to computational limitations, these works only perform short-term simulations, and it is not clear how the history of ejections by Didymos would take place in a long-term period. Since the YORP effect is continuously increasing Didymos rotation, it is possible that new avalanches will occur on the object, possibly re-injecting into the ring part of the material fallen onto Didymos \citep{trogolo2021}. In this case, we expect the total amount of material deposited in the ring to be greater than the amount of material actually detached from Didymos, since part of the material would be recycled. At the end of this process, Dimorphos and possible smaller objects are obtained, while the rest falls back onto Didymos, giving rise to its equatorial bulge \citep{hyodo2022}.

The viscous spreading is the consequence of inter-particle interactions within the ring (viscosity), where mass in transferred radially in order to conserve the total angular momentum of the system. We can associate three different viscosities to the viscous spreading: translational, collisional, and gravitational viscosity. Translational viscosity is related to the transfer of particle and is the result of momentum transport due to the random motion of particles \citep{Goldreich1978}. Collisional viscosity is an effect of momentum transfer due to impacts, where angular momentum is transported between the centre of the particles \citep{Araki1986}. Finally, the gravitational viscosity results from the scattering of particles due to self-gravity wakes structures in the ring \citep{Daisaka2001}. We include the three components of viscous spreading in our numerical simulations \citep{salmon2010}.

The ring spreading timescale ($\tau_{\rm vis}$) is a key parameter of the model, as it defines the formation pathways for Dimorphos, discussed in Section~\ref{sec:grow}, and the validity of the scenario envisioned by us. $\tau_{\rm YORP}$ gives the timescale for Didymos to reach the critical spin rate again after an avalanche, so we have that the material deposition will only be possible if $\tau_{\rm vis}$ be less than such a value. Otherwise, Didymos will push outward all the material overtime, preventing the formation of the bulge, which it is not in agreement with the numerical results obtained in \cite{hyodo2022}.

The YORP timescale is given by \citep{rubincam2000,Vokrouhlicky2002}
\begin{equation}
\tau_{\rm YORP}\sim\frac{3\pi c M_D}{5\Phi R_D^3T_D}(\alpha_D^2+\beta_D^2), 
\end{equation}
where $c$ is the speed of light and $\Phi$ is the solar flux the distance of Didymos distance from the Sun. We obtain $\tau_{\rm YORP}\sim10^5$~yr for Didymos.

In turn, the viscous spreading timescale can be written as \citep{Brahic1977,salmon2010}:
\begin{equation}
\tau_{\rm vis}=\frac{\zeta\Omega^3}{\Sigma^2}(a_{\rm FRL}-R_D)^2
\end{equation}
where $\Omega$ is the keplerian angular velocity, $a_{\rm FRL}$ is the location of the fluid Roche limit (FRL), $\Sigma$ the ring surface density, and $\zeta$ a parameter that depends on the location and physical parameters of the ring particles \citep{salmon2010}.

The surface density depends on the mass deposited in the ring, implying that the viscous spreading timescale will be intrinsically related to the timescale on which Didymos deposits material in the ring. For simplicity, we assume the instantaneous mass flux into the ring as 
\begin{equation}
\dot{M}(t)=\frac{M_T}{\tau}{\rm exp}\left(-\frac{t}{\tau}\right),
\end{equation}
where $M_T$ is the total mass deposited in the ring, $t$ is the simulation time, and $\tau$ is the deposition timescale. The total mass deposited in the ring at time $t$ is $M(t)=\int_0^{t'} \dot{M}(t')dt'$ ($M(\infty)=M_t$). The ring mass at every instant is much smaller than $M_T$ for cases with $\tau\neq0$.

We vary $M_T$ and $\tau$ in the ranges [$0.01-0.5$]~$M_D$ and [$0.1$-$10^4$]~yr, respectively. For such values, we obtain that the surface density vary in the range $\Sigma\sim10^2-10^4~kg/m^2$ when the satellites are formed, implying in $\tau_{\rm vis}\sim10-10^3$~yr, a value at least two orders of magnitude smaller than $\tau_{\rm YORP}$. A more robust approach to the mass deposited in the ring would require the study of avalanches at Didymos, depending on the its internal mass distribution and strength \citep[see][]{Zhang2017,Zhang2018,Zhang2021}, which is beyond the scope of this work. Here, our intention is to focus on processes related to the ring evolution.

For this, we carried out numerical simulations using the one-dimensional hybrid code \texttt{HYDRORINGS} \citep{charnoz2010,salmon2010}. The code couples two distinct modules: a finite volume module that tracks the viscous evolution of the ring  and satellite formation and an analytical orbital module to track the satellite evolution due to tidal effects and ring-satellite torques \citep[see][]{charnoz2011}. The code uses the formalism described in \cite{salmon2010} for modelling the ring's effective viscosity. Material that spreads beyond the fluid Roche limit ($a_{\rm FRL}=1.6926R_D$) is converted into one satellite per grid cell at each time-step. Collisions are assumed to happen when the distance between two satellites is less than twice their mutual Hill radius and are treated as perfect merge events. The formalism described in \cite{meyer1987} is used to account for satellite migration due to ring interaction. The code does not take into account mutual gravitational perturbations between the satellites. 

The ring is modelled as 50 cells distributed from the Didymos mean radius ($R_D$) until the FRL, composed of particles of size s=1~m with density $\rho_D$. Such a radius is roughly consistent in order of magnitude with the size of the constituent particles of Dimorphos, recorded by the DART mission. We simulate a scenario where Didymos starts with a disk of material orbiting inside its Roche Limit (case $\tau=0$), mimicking a scenario with a very fast deposition of material or with a debris disk due to an external impact \citep{Michel2020}. We also simulate a scenario where the ring region is initially empty. In such, we assume that Didymos deposited mass appears in the first cells of the grid (i.e. close to Didymos' surface). We perform simulations assuming $M_T=0.01$, $0.1$, $0.25$, and $0.5~M_D$ and $\tau=0.1$, $1$, $10^2$, and $10^4$~yr. 

We consider cases with and without tidal migration ($k_2/Q=0$, $\tau_{\rm tide}=0$). In the cases with tidal migration, we vary the tidal parameter $k_2/Q$ for Didymos within the value range expected for rubble-pile asteroids \citep{nimmo2019}: $k_2/Q=10^{-6}$, $10^{-5}$, $10^{-4}$, and $10^{-3}$, which correspond to tidal migration timescales $\tau_{\rm tide}\sim 10^3$, $10^2$, $10$, and $1$~yr, respectively. Such values of $\tau_{\rm tide}$ are at least an order of magnitude smaller than the BYORP timescale ($\sim 10^5$~yr, see Section~\ref{sec_parameters}) and as a simplification we do not include this effect in our simulations. Didymos is assumed to be a spherical object and we evolve the system until the ring surface density become less than the threshold value $10^{-1}$~kg/m$^2$. For such a value, the mass in a bin becomes less than the mass of a ring particle, condition for which we consider the ring region to be empty.

\section{Didymos starting with a ring} \label{sec:ring}
\begin{figure}[]
\subfloat[]{\includegraphics[width=0.49\columnwidth]{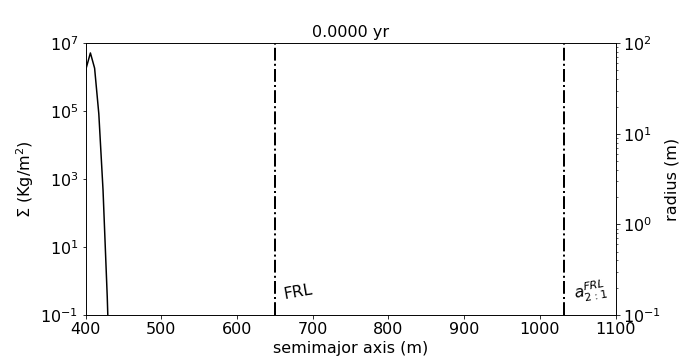}}
\subfloat[]{\includegraphics[width=0.49\columnwidth]{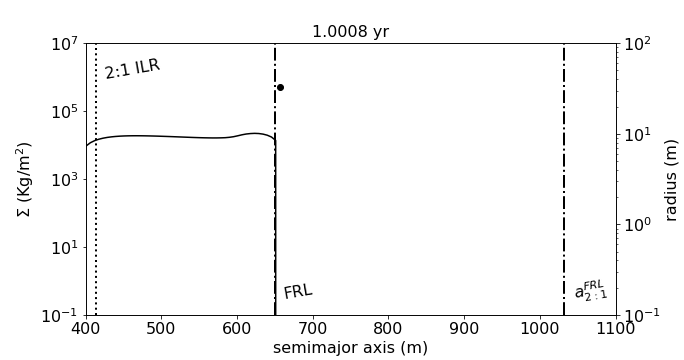}}
\\
\subfloat[]{\includegraphics[width=0.48\columnwidth]{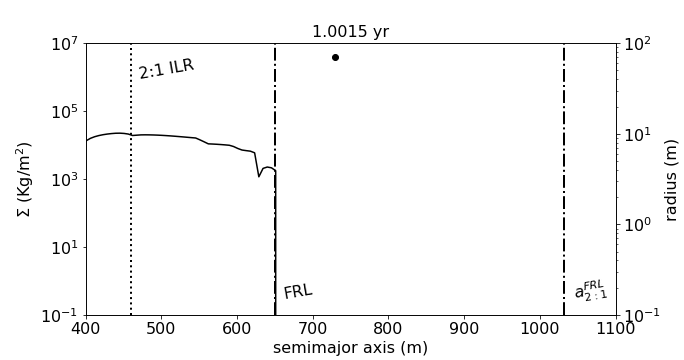}}
\subfloat[]{\includegraphics[width=0.48\columnwidth]{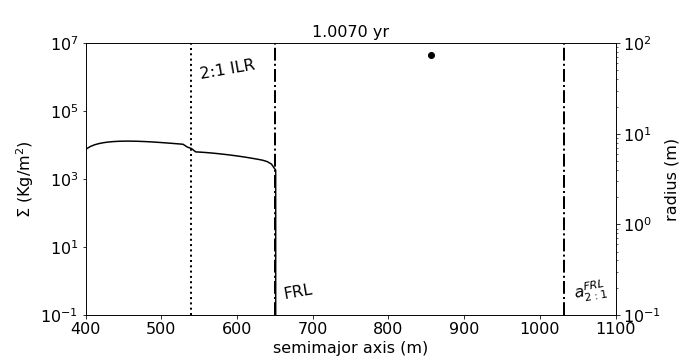}}
\\
\subfloat[]{\includegraphics[width=0.48\columnwidth]{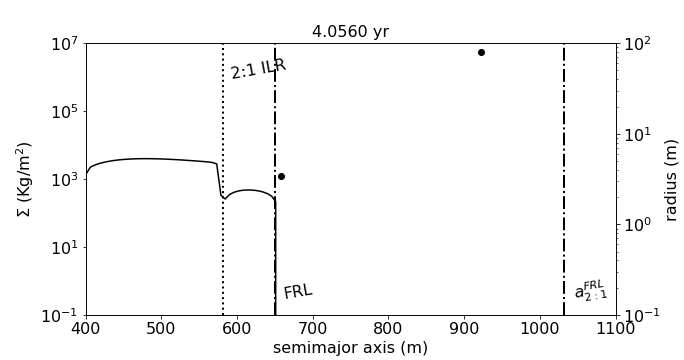}}
\subfloat[]{\includegraphics[width=0.48\columnwidth]{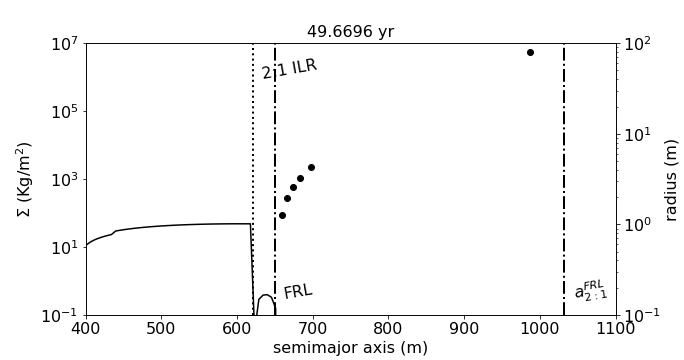}}
\\
\subfloat[]{\includegraphics[width=0.48\columnwidth]{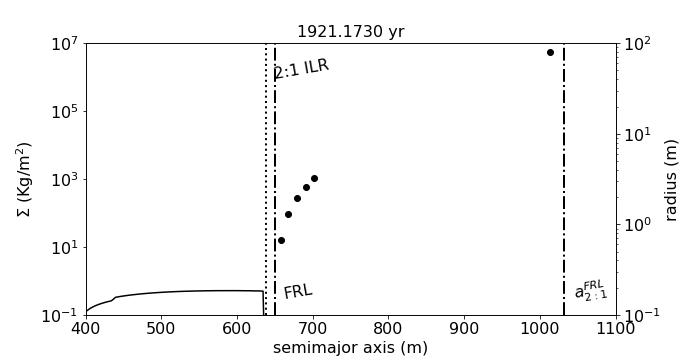}}
\subfloat[]{\includegraphics[width=0.48\columnwidth]{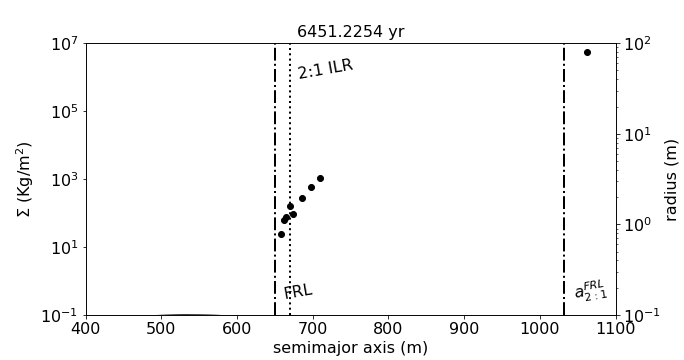}}
\\
\caption{Ring surface density (solid line, left scale) and satellite radius (black dots, right scale) as a function of distance to Didymos, for a system with $k_2/Q=10^{-5}$, starting with a ring of mass $M_T=0.25~M_D$. Each panel corresponds to a different snapshot where the simulation time is given at the top of the panel. The dashed lines give the Fluid Roche limit (FRL) and the maximum distance that satellites can migrate due to ring torques ($a_{2:1}^{FRL}$). The dotted line places the 2:1 inner Lindblad resonance location with Dimorphos (2:1 ILR). \label{fig:standard}}
\end{figure}

Once we described our dynamic model and numerical code, we will show the evolution of a case with tidal parameter $k_2/Q=10^{-5}$, starting with an initial ring of mass $M_T=0.25~M_D$ but without considering deposition of material (case with $\tau=0$). The different panels of Figure~\ref{fig:standard} show for different times (at the top of each panel), the ring surface density ($\Sigma$, solid line) and satellite radius (black dots) as a function of the distance to Didymos centre. The surface density is given on the left scale and the satellite radius on the right. 

The ring material spreads radially due to viscous effects (Fig.~\ref{fig:standard}a), with part of the material falling back onto Didymos and part flowing outwards, eventually crossing the FRL. Due to gravitational instabilities, material outside the FRL gives rise to a proto-Dimorphos (Fig.~\ref{fig:standard}b) that accretes all the ring material inside its Hill sphere \citep{karjalainen2007,charnoz2010,charnoz2011,hyodo2014}. When the satellite's mass increases, its Hill sphere grows, increasing the accretion. As a consequence, we observe a rapid growth of Dimorphos (Fig.~\ref{fig:standard}c) that reaches 0.93 Dimorphos masses in 0.01~yr ($<40~T_D$) after its formation  (Fig.~\ref{fig:standard}d).  This growth corresponds to a growth in the ``continuous regime'' according the classification of \cite{crida2012}. This means that the satellite grows progressively, at the Roche Limit, by accreting small pieces of ring material that are captured at the satellite surface. The satellite's critical Hill density is $\rho_c=M_D/(1.59a^3)\sim1.2~{\rm g/cm^3}$ \citep{porco2007}, which corresponds to a minimum density for a rubble-pile formed by the accumulation of material. Taking this value as fiducial, we obtain that the object is expected to have a maximum porosity of $\sim 50\%$.

While growing by accretion, Dimorphos migrates outward due to tidal migration and resonant interactions with the ring. When its Hill sphere no longer overlaps the ring, new moons can form from the ring, as can be seen in Fig.~\ref{fig:standard}e. In $t=4$~yr, the satellite is massive enough to confine part of the ring due to 2:1 inner Lindblad resonance (ILR, dotted line), which can be noticed by the step structure at about $570$~m. From this time on (Fig.~\ref{fig:standard}e-h), Dimorphos grows due to impacts with the newly formed moons \citep[``discrete regime'',][]{crida2012}. In about 40 years, the satellite mass reaches $0.98$ Dimorphos masses. 

In the following years of simulation, the almost formed Dimorphos confines part of the ring due to the 2:1 ILR (Fig.~\ref{fig:standard}f,g). Ring material mostly falls back onto Didymos. The unconfined material gives rise to objects with radius of meters. These objects do no migrate due to the weak ring torque. They accumulate near the Roche Limit. After $\sim 6400$~years, the ring is empty, leaving a population of debris around the FRL (Fig.~\ref{fig:standard}h). Work on the dynamics around non-uniformly shaped asteroids \citep{Scheeres2007,Madeira2022a,agruda2022,Ferrari2022} shows that the region near its surface is essentially unstable or chaotic. The irregularities in the asteroid induce variations in the eccentricity of nearby objects, that will have limited lifetimes, colliding or being ejected from the system. This could be the fate of the debris population, given its proximity to Didymos surface. In the next sections, we discuss how our results change when we consider the scenario with material deposition by Didymos.

\section{On the flow and tidal parameters} \label{sec_parameters} 
\begin{figure}[]
\centering
\subfloat[]{\includegraphics[width=0.7\columnwidth]{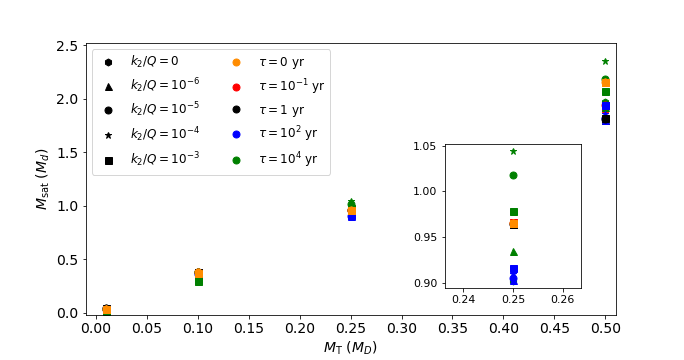}}
\\
\subfloat[]{\includegraphics[width=0.7\columnwidth]{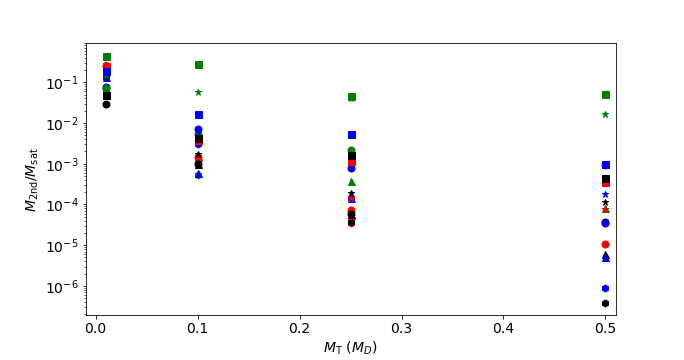}}
\caption{a) Mass of the largest satellite (in Dimorphos masses) and b) mass ratio between second largest and largest object on the system as a function of the system parameters. The total mass deposited in the ring is given on the x-axis. The different colours stand to the different deposition timescales and the markers correspond to the tidal parameters. \label{fig:parameters}}
\end{figure}

Figure~\ref{fig:parameters} summarises all our numerical simulations, showing the mass of the largest satellite and mass ratio between second largest and largest object formed in the simulation as a function of the total mass deposited in the ring (x-axis), deposition timescale (colours), and tidal parameter (markers). We find that the mass of the largest satellite is mostly controlled by the total mass deposited in the ring ($M_T$), regardless of whether the mass is delivered at the very beginning of the simulation or not (Fig~\ref{fig:parameters}a), meaning that $\tau$ and $k_2/Q$ have only little influence on Dimorphos mass. The mass of the largest satellite is directly proportional to the total mass deposited in the ring. We obtain that a mass of $0.25$ Didymos masses is required to form a satellite with the mass of Dimorphos (smaller panel in Fig~\ref{fig:parameters}a).

We clarify once again that $M_T$ corresponds to the total amount of material deposited in the ring over time and not to the mass of the ring at any instant \citep[as in works like][]{crida2012,charnoz2011,madeira2022}. We get that the ring will have its maximum mass just before the formation of proto-Dimorphos, reaching a value up to $0.1~M_D$ in the simulations with $M_T=0.25~M_D$. If we assume that Didymos can re-inject part of the fallen material into the ring, we have that the value $0.1~M_D$ can be interpreted as a lower bound for the mass detached from Didymos in a single avalanche.

\cite{hyodo2022} find that a mass of the order of $10\%$ of the primary is ejected via a single landslide when the resultant shape is a top-shape \citep[cases with spin-up timescales $\lesssim$ few days and effective friction angles $\gtrsim 70^{\circ}$,][]{sugiura2021}. They obtain that the timescale of the mass ejection (i.e., landslide) is comparable to the critical spin period, that is, $\sim 2-3$h (i.e., $\sim 10^4$ seconds). This corresponds to $\tau \sim 10^{-4}$~yr in our numerical description of mass deposition, implying that the formation of Dimorphos may be somewhat similar to the scenario shown in Figure~\ref{fig:standard}. 

The correlation between the ring mass and the largest satellite mass is quantitatively demonstrated by 3D direct $N$-body simulations of ring spreading \citep{hyodo2015}, and was also found in \cite{charnoz2010,charnoz2011}. We point out that the mass of the largest satellite depends on the initial angular momentum of the ring \citep{hyodo2015} -- consequently, it depends on the initial width of the ring. If the ring is initially wider, as would be expected in a scenario with a ring formed due to an impact, the mass needed to form Dimorphos becomes smaller. Therefore, the numbers obtained by us should not be taken as engraved on marble, but as responsible for assigning orders of magnitude to the system quantities.

We also get a relation between $M_T$ and the mass ratio of the second largest satellite (Fig~\ref{fig:parameters}b). When the largest satellite is very massive, it accretes a large amount of ring material. As a consequence, only small objects will form in the system. Systems with larger $M_T$ form a very massive satellite and some meter-sized bodies, while systems with smaller $M_T$ forms dozens of objects with similar mass ratio -- with radius of tens of meters. These objects are in the ''pyramidal regime'' defined by \cite{crida2012}. The exceptions are the cases with $\tau=10^4$~yr and $k_2/Q=10^{-3}$, in which the most massive satellite migrates fast enough to allow other massive satellites to form \citep{charnoz2011}.

The deposition timescale mainly controls the formation timescale. In general, the cases with $\tau=10^4$~yr form Dimorphos that are slightly more massive than those obtained in simulations with $\tau<10^4$~yr. Cases with $\tau=10^2$~yr, however, form Dimorphos that are less massive than those obtained in simulations with $\tau<1$~yr, for most cases. Despite the weak dependency of Dimorphos mass on deposition timescale, we get different formation pathways depending on $\tau$, as will be discussed in Section~\ref{sec:grow}.

We also find no clear relation between the tidal parameter and the mass of the largest satellite. However, we do find an effect of tidal parameter on the mass of the second largest object. The resonant interactions between satellite and ring is responsible for both the outward migration of satellite and the radial confinement of the ring, with the 2:1 ILR location with the FRL ($a_{2:1}^{FRL}=1049$~m) being the maximum location to which a satellite can migrate due to this effect \citep{charnoz2011,madeira2022}. Upon reaching this position, the satellite confines most of the ring material and only tiny satellites can be formed, as observed in the cases with $k_2/Q=0$. In such simulations, Dimorphos does not migrate after reaching $a_{2:1}^{FRL}$. Due to the tidal migration, however, the satellite eventually migrates out of the $a_{2:1}^{FRL}$ position, leaving the ring  unconfined. Thus, larger objects can form. A larger $k_2/Q$ translates to shorter periods of ring confinement, which explains why larger objects are obtained in systems with larger $k_2/Q$.
\begin{figure}[]
\centering
\includegraphics[width=0.7\linewidth]{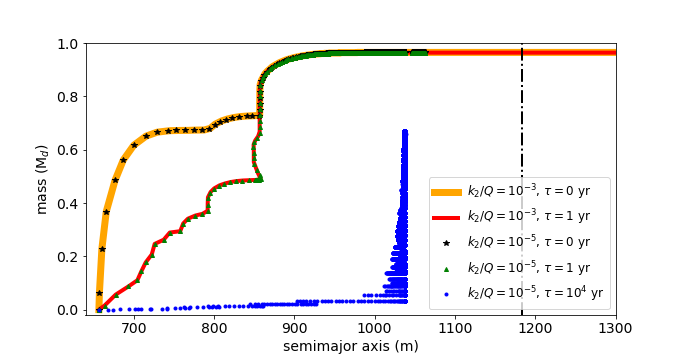}
\caption{Evolution of Dimorphos mass as a function of the semimajor axis for different simulations with $M_T=0.25~M_D$. Each colour corresponds to a different simulation. The solid lines correspond to cases with $k_2/Q=10^{-3}$ and the markers to cases with $k_2/Q=10^{-5}$. The vertical dashed line gives the current location of Dimorphos. \label{fig:mass_sma}}
\end{figure}

As expected, the parameter $k_2/Q$ will control the location of the satellite when the ring disappears. Figure~\ref{fig:mass_sma} shows the evolution of mass and semimajor axis of Dimorphos in simulations with $M_T=0.25~M_D$. As can be seen in all the simulations with $k_2/Q=10^{-3}$ (solid line), the ring disappears when the satellite is out of the current location of Dimorphos (vertical dashed line). In fact, we find that the ring is completely emptied before Dimorphos reaches its current location only for $k_2/Q\leq10^{-5}$. These results lead us to conclude that the Didymos tidal parameter must be in this range $k_2/Q\leq10^{-5}$. 

The orbit of Dimorphos may be further affected after its formation by other physical processes, such as the binary YORP (BYORP) effect \citep[e.g.,][]{cuk2005}. Such an effect is an extension of the YORP effect, affecting a binary system when the satellite is in a spin-orbit resonance (as in the case of Dimorphos). Due to the BYORP, the satellite's orbit can be expanded or contracted, depending on the physical properties of the satellite, which are expressed mathematically through the dimensionless coefficient $B_s$ \citep{cuk2005}. \cite{hyodo2022} propose that the top-shaped asteroids, such as Ryugu and Bennu, could have at least one satellite in the past, with the BYORP effect being the responsible for expanding the orbits of these objects on a relatively short timescale of $\sim 10^5$~yr, until their eventual ejection from the system.

By analogy, we could speculate that a similar process is happening in the Didymos system, that is, Dimorphos is recently formed and is now on the way to being ejected. However, as BYORP and tidal effects can act in opposite directions, an equilibrium state can be reached for \citep{walsh2015}:
\begin{equation}
B_s=\frac{4\pi^2}{3}\frac{k_2}{Q}\left(\frac{M_d}{M_D}\right)^{4/3}\frac{\rho_D^2R_D^2G}{\Phi a_d^7}.    
\end{equation}
For the tidal parameter $k_2/Q=10^{-5}$, we get $B_s=0.0064$ which is in line with the expected values for satellites of top-shaped asteroids \citep[$B_s\sim 0.01$,][]{McMahon2010,walsh2015}. Therefore, it is possible that Dimorphos formed a long time ago and has kept being at an equilibrium location around Didymos. Next, we explore the effects of $\tau$ on the formation of Dimorphos.

\section{Formation pathways for Dimorphos} \label{sec:grow}
\begin{figure}[]
\centering
\subfloat[]{\includegraphics[width=0.8\columnwidth]{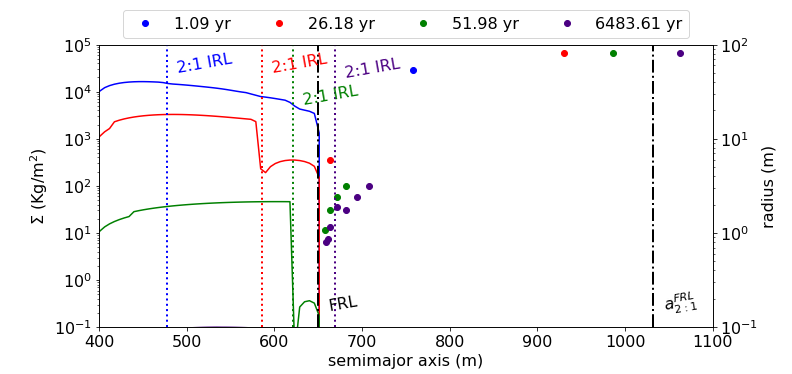}}\\
\subfloat[]{\includegraphics[width=0.8\columnwidth]{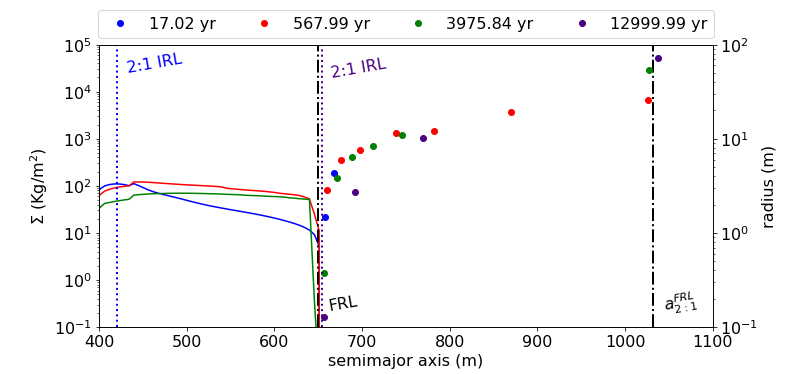}}
\caption{Left scale of panels shows ring surface density (solid line) and right scale, the radius of the satellites (black dots) for systems with $M_T=0.25~M_D$ and $k_2/Q=10^{-5}$. The top and bottom panel correspond to the cases with $\tau=1$~yr and $10^4$~yr, respectively. The different colors correspond to different times (shown at the top of each panel). The Fluid Roche limit (FRL) and the maximum distance that satellites can migrate due to ring torques ($a_{2:1}^{FRL}$) are given by the dashed lines, while the 2:1 ILR location with Dimorphos is given by the dotted lines. \label{fig:compare}}
\end{figure}

Different formation pathways are obtained for Dimorphos, depending on the deposition timescale. For $\tau=0$, the pathway is the one described in Section~\ref{sec:ring}: a single satellite forms and reaches a mass $>0.9~M_d$ in the continuous regime. When this is far enough away from the FRL, smaller moons begin to form. The satellite then starts to grow through impacts with these moons (discrete regime).

When we increase the deposition time, we decrease the mass flux and the amount of mass available to build the first satellite is smaller. So, we get a smaller initial satellite. The Hill radius of the satellite will also be smaller and, as a consequence, it will accrete less material directly from the ring. This can be seen in the difference of slope of the curves with $\tau=0$ and $1$~yr in Figure~\ref{fig:mass_sma}. Figure~\ref{fig:compare} shows the ring surface density (solid line, left scale) and satellite radius (black dots, right scale) of simulations with $M_T=0.25~M_D$ and $k_2/Q=10^{-5}$. The top panel is the case with $\tau=1$~yr and the bottom one is the case with $\tau=10^4$~yr. The different colours correspond to the system at the different times indicated at the top of each panel.

For $\tau=1$~yr, the system starts without ring but receives material ejected by Didymos, which spreads out of the FRL in $1.09$~yr. A small satellite forms at the Roche limit. The ring is supplied by Didymos, which in turn feeds the satellite that reaches $0.3~M_d$ in 0.01~yr after its formation. Then, new satellites form and Dimorphos grows to its current mass by accreting meter-sized satellites in the discrete regime. For high deposition rates, we have a rapid formation of Dimorphos, first by accretion of ring material and then by impacts with smaller moons, and the parameter $\tau$ is responsible for defining the amount of material accreted by Dimorphos at each stage. The accretion of Dimorphos directly from ring material is called ``continuous regime'' according the classification of \cite{crida2012}.

Now, for $\tau=10^2$ and $10^4$~yr, satellite formation takes place in a different way. Due to the small mass flux in the ring, meter-sized moons form and migrate outwards due to ring torque, allowing the formation of new moons of tens of meters in radius (Fig.~\ref{fig:compare}, right column). Therefore, the system evolves into a collection of moons as seen in the right column of Figure~\ref{fig:compare}, with Dimorphos being formed by merge events between these bodies. According to the classification of \cite{crida2012}, this corresponds to complete formation in the pyramidal regime, which would imply Dimorphos with different properties than those obtained in the case of the continuous accretion regime (see Section~\ref{subsec:shape}).

\section{Discussion} \label{sec_discussion}
\subsection{Dimorphos Shape} \label{subsec:shape}
Whereas we do not directly track the shape of Dimorphos in our simulation, previous studies have shown that different accreting regimes may lead to different shape. Here, in light of our results, we build assumptions for the possible shape of Dimorphos based on the different accretion regimes in which the satellite can be formed. The influence of the formation environment on the shape of a satellite is an intricate problem and it is not our intention to draw conclusions about Dimorphos physical properties, but only to raise hypotheses for future work.

Although the Dimorphos mass is mostly controlled by the total mass deposited in the ring ($M_T$), we obtain completely different formation pathways depending on the deposition timescale $\tau$. For short timescales ($\tau\lesssim$yr), Dimorphos grows in both continuous and discrete regimes, while for large timescales, all growth occurs in the pyramidal regime. These formation pathways are expected to affect the final shape and structure of Dimorphos.

Material flowing out of the FRL collapses into seeds \citep{karjalainen2007} which accrete the ring material inside their Hill sphere. They could form as lemon-shaped self-gravitating aggregates \citep{porco2007,Tiscareno2013}. The maximum porosity expected for these objects in the Didymos system is $50\%$. For high deposition rate, Dimorphos accretes a fraction of its mass in this way and then starts to grow by impacts with smaller satellites. Due to the low impact velocities, smaller satellites are expected to sediment onto Dimorphos, preserving its general porosity \citep{benz2000,Jutzi2015}. Depending on the impact geometry (grazing impacts), a significant amount of debris can be released, causing deposition on the crater and also restructuring due to large deformations \citep{Jutzi2008a,Jutzi2009b}. Such processes favour the formation of a smooth regions on Dimorphos surface, preserving (at least in part) the ellipsoidal shape of the object. This scenario seems to be compatible with the Dimorphos images obtained by DART mission.

For low deposition rates, Dimorphos would be the result of impacts between gravitational aggregates of similar sizes. As the objects are in nearly circular and coplanar orbits, impacts are expected to occur at low velocities, resulting in a merge or partial accretion event, depending on the impact geometry \citep{Leinhardt2012}. \cite{Leleu2018} show that, depending on the collisions conditions, the impact of two moons can also lead to hit and run events before they finally merge. \citeauthor{Leleu2018} find that growth in the pyramidal regime can produce flattened satellites with ridges or objects with elongate irregular shapes. Impacts in the Didymos system are not expected to reduce porosity, however mechanisms such as shear dilatation or re-accumulation can increase porosity. Thus, Dimorphos may have higher porosity than their parent bodies. Regarding the composition, Dimorphos is expected to have a similar composition to the surface of Didymos, source of the ring material.

\subsection{Solar radiation pressure and modelling limitations}
Due to their proximity to the Sun, NEAs are strongly affected by angular momentum exchanges with the solar radiation \citep{walsh2015,margot2015}. A classic example of this is the YORP effect (Section~\ref{sec_dynamical}), responsible for increasing the asteroid's spin. The particles around the NEAs are also expected to be affected by the different components of the solar radiation force, such as radiation pressure, Poynting-Robertson drag, and Yarkovsky effects \citep{Burns1979}. The last two effects are responsible for variation in the semi-major axis and spin rate on timescales of $\gtrsim10^6$~yr for our system \citep{Mignard1984,rubincam2000}. Thus, they are not expected to significantly affect the formation of Dimorphos. However, it is possible that the ring particles are affected by radiation pressure.

Radiation pressure is caused by the transfer of momentum due to impacts of solar radiation on the ring particles, being responsible for causing variations in eccentricity in periods of few orbits \citep{Hamilton1993} and small periodic variations in the semi-major axis \citep{Madeira2018,Madeira2020}. The variation in eccentricity due to radiation pressure can be estimated as \citep{Hamilton1996} 
\begin{equation}
e_{\rm RP}=\frac{C}{\sqrt{1+C^2}}    
\end{equation}
where \citep{Hamilton1996}
\begin{equation}
C=\frac{9}{8}\frac{Q_{\rm pr}\Phi}{nac\rho_Ds}    
\end{equation}
in which $n$ is the orbital frequency and $Q_{\rm pr}$ is the solar radiation efficiency, computed from Mie theory \citep{Irvine1965}.

For one-meter particles made of ice we get $e_{\rm RP}\sim 10^{-3}$, a value small enough that the particles do not collide with the central primary at its pericenter distance, so in this case radiation pressure can be negligible. Nonetheless, such an increase in eccentricity is responsible for increasing the velocity of collisions between particles, which may result in fragmentation and grinding into smaller particles. This means that part of the ring material might be lost in some orbits due to this effect. Now, if we assume that the particles are a mixture of ice with a small fraction of silicates, we have $e_{\rm RP}\sim 10^{-4}$, dropping to values of the order of $10^{-5}$ for particles made of more resistant materials or ice particles with a silicate concentration of dozens of percent \citep{Irvine1965,Artymowicz1988}, which will probably be the case for particles ejected by Didymos. For such cases, we expect the radiation pressure to have a minor effect on Dimorphos formation for the particle size assumed in our simulations (one meter).

We point out that particles with radii of the order of mm to cm must also compose the ring, as data on NEAs indicate the possible presence of material with this size range on the surface of Didymos \citep{pajola2022}. Such particles will present eccentricities $\gtrsim0.01$, therefore, it is expected that the evolution of the system will be more complex when considering their presence in the disk. Here, we consider the ring composed only of meter-sized particles due to a computational limitation.  

More sophisticated simulations of the ring evolution with solar radiation and fragmentation would require tracking the evolution of the particles as single entities and with more than one size in a same simulation, which is not possible with the current version of our code. The \texttt{HYDRORINGS} code describes a ring as a 1D entity and using a hydrodynamic approach, for which there is still no formalism with multiple particle sizes in a closed form. Furthermore, a more realistic study of the formation of Dimorphos from a ring requires the investigation of different size distributions in the ring. We leave this investigation for future work.

Another effect that likely removes material from the ring is Didymos shape irregularities. As demonstrated by a set of papers \citep[eg.][]{Scheeres2007,Madeira2022a,agruda2022,Ferrari2022}, the region in the vicinity of non-uniformly shaped objects is normally unstable (or chaotic) and the particles have limited lifetimes. These are two caveats of our model. Including the actual shape of Didymos in our simulations would make it impossible to track the ring's evolution over billions of orbits with current computer capacities. This is the same limitation presented by works that reproduce Didymos shape, such as \cite{Yu2018,Zhang2017,Zhang2021} and \cite{hyodo2022}, for which the system can be evolved only by a limited number of orbits. The results of this article should be treat as a first-order study (like other studies of this kind), and the ring masses as lower bounds, since material can be removed due to Didymos shape and solar radiation, among others unmentioned external effects.

\section{Conclusion} \label{sec_conclusion}
In this work, we studied the formation of Dimorphos from material released from Didymos. The YORP effect increases the spinning of Didymos, detaching surface material that settles in a ring around the primary. Most of the material should falls back onto Didymos, possibly being re-injected in the ring due to Didymos fast spin, while the remainder would give rise to satellite seeds outside the Roche limit, and finally to Dimorphos. Through numerical simulations, we analysed the evolution of the ring material, assuming an e-folding temporal function for the mass deposited in the ring. We varied the total mass deposited in the ring, deposition timescale, and Didymos tidal parameter. 

Dimorphos with its current mass is obtained for a total mass deposited in the ring equal to $25\%$ of Didymos' mass, regardless of the timescale on which the material is delivered. The material falling on Didymos is possibly responsible for giving birth to the asteroid's equatorial ridge \citep{hyodo2022}. Some of the material can also be re-injected into the ring due to Didymos spin-up. Given the simplicity of our model, we believe that our results can be applied qualitatively to all similar binary systems with a primary close to the critical rotation. We leave this work for a future publication.

Dimorphos is formed in its current position without a ring for the a Didymos tidal parameter of $k_2/Q\leq10^{-5}$, being obtained that the satellite can be kept in its position through a BYORP-tidal balance for $k_2/Q\sim10^{-5}$. Different deposition timescales can lead to different shapes of Dimorphos. For example, timescales $\geq10^2$~yrs would give rise to Dimorphos with an irregular shape due to accretion in pyramidal regime. However, the deposition timescale is expected to be less than one year \citep{sugiura2021}, which would give rise to a nearly lemon-shaped (ellipsoidal) Dimorphos composed of meter sized rocks, which seems to be in agreement with recent images obtained by DART mission. DART and HERA spacecraft will allow us to obtain details on the shapes and composition of Didymos and Dimorphos and improve models of formation of binary systems.

\section*{Acknowledgments}
G.M. thanks FAPESP for financial support via grants 2018/23568-6 and 2021/07181-7. R.H. acknowledges the financial support of MEXT/JSPS KAKENHI (Grant Number JP22K14091). R.H. also acknowledges JAXA's International Top Young program. We acknowledge Dr. Keisuke Sugiura for discussion. Thanks to the reviewers for the comments that helped us to improve the article.

\bibliography{references}  

\begin{thebibliography}{71}
\providecommand{\natexlab}[1]{#1}
\providecommand{\url}[1]{\texttt{#1}}
\expandafter\ifx\csname urlstyle\endcsname\relax
  \providecommand{\doi}[1]{doi: #1}\else
  \providecommand{\doi}{doi: \begingroup \urlstyle{rm}\Url}\fi

\bibitem[{Agrusa} et~al.(2022){Agrusa}, {Ferrari}, {Zhang}, {Richardson}, and
  {Michel}]{agruda2022}
H.~F. {Agrusa}, F.~{Ferrari}, Y.~{Zhang}, D.~C. {Richardson}, and P.~{Michel}.
\newblock {Dynamical Evolution of the Didymos-Dimorphos Binary Asteroid as
  Rubble Piles following the DART Impact}.
\newblock \emph{The Planetary Science Journal}, 3\penalty0 (7):\penalty0 158,
  July 2022.
\newblock \doi{10.3847/PSJ/ac76c1}.

\bibitem[{Araki} and {Tremaine}(1986)]{Araki1986}
S.~{Araki} and S.~{Tremaine}.
\newblock {The dynamics of dense particle disks}.
\newblock \emph{Icarus}, 65\penalty0 (1):\penalty0 83--109, Jan. 1986.
\newblock \doi{10.1016/0019-1035(86)90065-5}.

\bibitem[{Artymowicz}(1988)]{Artymowicz1988}
P.~{Artymowicz}.
\newblock {Radiation Pressure Forces on Particles in the Beta Pictoris System}.
\newblock \emph{Astrophysical Journal Letters}, 335:\penalty0 L79, Dec. 1988.
\newblock \doi{10.1086/185344}.

\bibitem[{Becker} et~al.(2015){Becker}, {Howell}, {Nolan}, {Magri}, {Pravec},
  {Taylor}, {Oey}, {Higgins}, {Vil{\'a}gi}, {Korno{\v{s}}}, {Gal{\'a}d},
  {Gajdo{\v{s}}}, {Gaftonyuk}, {Krugly}, {Molotov}, {Hicks}, {Carbognani},
  {Warner}, {Vachier}, {Marchis}, and {Pollock}]{becker2015}
T.~M. {Becker}, E.~S. {Howell}, M.~C. {Nolan}, C.~{Magri}, P.~{Pravec}, P.~A.
  {Taylor}, J.~{Oey}, D.~{Higgins}, J.~{Vil{\'a}gi}, L.~{Korno{\v{s}}},
  A.~{Gal{\'a}d}, {\v{S}}.~{Gajdo{\v{s}}}, N.~M. {Gaftonyuk}, Y.~N. {Krugly},
  I.~E. {Molotov}, M.~D. {Hicks}, A.~{Carbognani}, B.~D. {Warner},
  F.~{Vachier}, F.~{Marchis}, and J.~T. {Pollock}.
\newblock {Physical modeling of triple near-Earth Asteroid (153591) 2001
  SN$_{263}$ from radar and optical light curve observations}.
\newblock \emph{Icarus}, 248:\penalty0 499--515, Mar. 2015.
\newblock \doi{10.1016/j.icarus.2014.10.048}.

\bibitem[Benz(2000)]{benz2000}
W.~Benz.
\newblock Low velocity collisions and the growth of planetesimals.
\newblock In \emph{From dust to terrestrial planets}, pages 279--294. Springer,
  2000.

\bibitem[{Bottke} et~al.(2006){Bottke}, {Vokrouhlick{\'y}}, {Rubincam}, and
  {Nesvorn{\'y}}]{Bottke2006}
J.~{Bottke}, William~F., D.~{Vokrouhlick{\'y}}, D.~P. {Rubincam}, and
  D.~{Nesvorn{\'y}}.
\newblock {The Yarkovsky and Yorp Effects: Implications for Asteroid Dynamics}.
\newblock \emph{Annual Review of Earth and Planetary Sciences}, 34:\penalty0
  157--191, May 2006.
\newblock \doi{10.1146/annurev.earth.34.031405.125154}.

\bibitem[{Brahic}(1977)]{Brahic1977}
A.~{Brahic}.
\newblock {Systems of colliding bodies in a gravitational field: I - Numerical
  simulations of the standard model.}
\newblock \emph{Astronomy and Astrophysics}, 54:\penalty0 895--907, Feb. 1977.

\bibitem[{Brozovi{\'c}} et~al.(2011){Brozovi{\'c}}, {Benner}, {Taylor},
  {Nolan}, {Howell}, {Magri}, {Scheeres}, {Giorgini}, {Pollock}, {Pravec},
  {Gal{\'a}d}, {Fang}, {Margot}, {Busch}, {Shepard}, {Reichart}, {Ivarsen},
  {Haislip}, {LaCluyze}, {Jao}, {Slade}, {Lawrence}, and {Hicks}]{Brozovic2011}
M.~{Brozovi{\'c}}, L.~A.~M. {Benner}, P.~A. {Taylor}, M.~C. {Nolan}, E.~S.
  {Howell}, C.~{Magri}, D.~J. {Scheeres}, J.~D. {Giorgini}, J.~T. {Pollock},
  P.~{Pravec}, A.~{Gal{\'a}d}, J.~{Fang}, J.-L. {Margot}, M.~W. {Busch}, M.~K.
  {Shepard}, D.~E. {Reichart}, K.~M. {Ivarsen}, J.~B. {Haislip}, A.~P.
  {LaCluyze}, J.~{Jao}, M.~A. {Slade}, K.~J. {Lawrence}, and M.~D. {Hicks}.
\newblock {Radar and optical observations and physical modeling of triple
  near-Earth Asteroid (136617) 1994 CC}.
\newblock \emph{Icarus}, 216\penalty0 (1):\penalty0 241--256, Nov. 2011.
\newblock \doi{10.1016/j.icarus.2011.09.002}.

\bibitem[{Burns} et~al.(1979){Burns}, {Lamy}, and {Soter}]{Burns1979}
J.~A. {Burns}, P.~L. {Lamy}, and S.~{Soter}.
\newblock {Radiation forces on small particles in the solar system}.
\newblock \emph{Icarus}, 40\penalty0 (1):\penalty0 1--48, Oct. 1979.
\newblock \doi{10.1016/0019-1035(79)90050-2}.

\bibitem[{Charnoz} et~al.(2010){Charnoz}, {Salmon}, and {Crida}]{charnoz2010}
S.~{Charnoz}, J.~{Salmon}, and A.~{Crida}.
\newblock {The recent formation of Saturn's moonlets from viscous spreading of
  the main rings}.
\newblock \emph{Nature}, 465\penalty0 (7299):\penalty0 752--754, June 2010.
\newblock \doi{10.1038/nature09096}.

\bibitem[{Charnoz} et~al.(2011){Charnoz}, {Crida}, {Castillo-Rogez}, {Lainey},
  {Dones}, {Karatekin}, {Tobie}, {Mathis}, {Le Poncin-Lafitte}, and
  {Salmon}]{charnoz2011}
S.~{Charnoz}, A.~{Crida}, J.~C. {Castillo-Rogez}, V.~{Lainey}, L.~{Dones},
  {\"O}.~{Karatekin}, G.~{Tobie}, S.~{Mathis}, C.~{Le Poncin-Lafitte}, and
  J.~{Salmon}.
\newblock {Accretion of Saturn{\textquoteright}s mid-sized moons during the
  viscous spreading of young massive rings: Solving the paradox of
  silicate-poor rings versus silicate-rich moons}.
\newblock \emph{Icarus}, 216\penalty0 (2):\penalty0 535--550, Dec. 2011.
\newblock \doi{10.1016/j.icarus.2011.09.017}.

\bibitem[{Cheng} et~al.(2018){Cheng}, {Rivkin}, {Michel}, {Atchison},
  {Barnouin}, {Benner}, {Chabot}, {Ernst}, {Fahnestock}, {Kueppers}, {Pravec},
  {Rainey}, {Richardson}, {Stickle}, and {Thomas}]{cheng2018}
A.~F. {Cheng}, A.~S. {Rivkin}, P.~{Michel}, J.~{Atchison}, O.~{Barnouin},
  L.~{Benner}, N.~L. {Chabot}, C.~{Ernst}, E.~G. {Fahnestock}, M.~{Kueppers},
  P.~{Pravec}, E.~{Rainey}, D.~C. {Richardson}, A.~M. {Stickle}, and
  C.~{Thomas}.
\newblock {AIDA DART asteroid deflection test: Planetary defense and science
  objectives}.
\newblock \emph{Planet.~Space~Sci.}, 157:\penalty0 104--115, Aug. 2018.
\newblock \doi{10.1016/j.pss.2018.02.015}.

\bibitem[{Crida} and {Charnoz}(2012)]{crida2012}
A.~{Crida} and S.~{Charnoz}.
\newblock {Formation of Regular Satellites from Ancient Massive Rings in the
  Solar System}.
\newblock \emph{Science}, 338\penalty0 (6111):\penalty0 1196, Nov. 2012.
\newblock \doi{10.1126/science.1226477}.

\bibitem[{{\'C}uk} and {Burns}(2005)]{cuk2005}
M.~{{\'C}uk} and J.~A. {Burns}.
\newblock {Effects of thermal radiation on the dynamics of binary NEAs}.
\newblock \emph{Icarus}, 176\penalty0 (2):\penalty0 418--431, Aug. 2005.
\newblock \doi{10.1016/j.icarus.2005.02.001}.

\bibitem[{Daisaka} et~al.(2001){Daisaka}, {Tanaka}, and {Ida}]{Daisaka2001}
H.~{Daisaka}, H.~{Tanaka}, and S.~{Ida}.
\newblock {Viscosity in a Dense Planetary Ring with Self-Gravitating
  Particles}.
\newblock \emph{Icarus}, 154\penalty0 (2):\penalty0 296--312, Dec. 2001.
\newblock \doi{10.1006/icar.2001.6716}.

\bibitem[{de Le{\'o}n} et~al.(2010){de Le{\'o}n}, {Licandro}, {Serra-Ricart},
  {Pinilla-Alonso}, and {Campins}]{deLeon2010}
J.~{de Le{\'o}n}, J.~{Licandro}, M.~{Serra-Ricart}, N.~{Pinilla-Alonso}, and
  H.~{Campins}.
\newblock {Observations, compositional, and physical characterization of
  near-Earth and Mars-crosser asteroids from a spectroscopic survey}.
\newblock \emph{A\&A}, 517:\penalty0 A23, July 2010.
\newblock \doi{10.1051/0004-6361/200913852}.

\bibitem[{Ferrari} et~al.(2022){Ferrari}, {Raducan}, {Soldini}, and
  {Jutzi}]{Ferrari2022}
F.~{Ferrari}, S.~D. {Raducan}, S.~{Soldini}, and M.~{Jutzi}.
\newblock {Ejecta Formation, Early Collisional Processes, and Dynamical
  Evolution after the DART Impact on Dimorphos}.
\newblock \emph{The Planetary Science Journal}, 3\penalty0 (7):\penalty0 177,
  July 2022.
\newblock \doi{10.3847/PSJ/ac7cf0}.

\bibitem[{Goldreich} and {Tremaine}(1978)]{Goldreich1978}
P.~{Goldreich} and S.~D. {Tremaine}.
\newblock {The velocity dispersion in Saturn's rings}.
\newblock \emph{Icarus}, 34\penalty0 (2):\penalty0 227--239, May 1978.
\newblock \doi{10.1016/0019-1035(78)90164-1}.

\bibitem[{Hamilton}(1993)]{Hamilton1993}
D.~P. {Hamilton}.
\newblock {Motion of Dust in a Planetary Magnetosphere: Orbit-Averaged
  Equations for Oblateness, Electromagnetic, and Radiation Forces with
  Application to Saturn's E Ring}.
\newblock \emph{Icarus}, 101\penalty0 (2):\penalty0 244--264, Feb. 1993.
\newblock \doi{10.1006/icar.1993.1022}.

\bibitem[{Hamilton} and {Krivov}(1996)]{Hamilton1996}
D.~P. {Hamilton} and A.~V. {Krivov}.
\newblock {Circumplanetary Dust Dynamics: Effects of Solar Gravity, Radiation
  Pressure, Planetary Oblateness, and Electromagnetism}.
\newblock \emph{Icarus}, 123\penalty0 (2):\penalty0 503--523, Oct. 1996.
\newblock \doi{10.1006/icar.1996.0175}.

\bibitem[{Hirabayashi} et~al.(2014){Hirabayashi}, {Scheeres}, {S{\'a}nchez},
  and {Gabriel}]{Hirabayashi2014}
M.~{Hirabayashi}, D.~J. {Scheeres}, D.~P. {S{\'a}nchez}, and T.~{Gabriel}.
\newblock {Constraints on the Physical Properties of Main Belt Comet P/2013 R3
  from its Breakup Event}.
\newblock \emph{ApJL}, 789\penalty0 (1):\penalty0 L12, July 2014.
\newblock \doi{10.1088/2041-8205/789/1/L12}.

\bibitem[{Hirabayashi} et~al.(2017){Hirabayashi}, {Schwartz}, {Yu}, {Davis},
  {Chesley}, {Fahnestock}, {Michel}, {Richardson}, {Naidu}, {Scheeres},
  {Cheng}, {Rivkin}, and {Benner}]{Hirabayashi2017}
M.~{Hirabayashi}, S.~R. {Schwartz}, Y.~{Yu}, A.~B. {Davis}, S.~R. {Chesley},
  E.~G. {Fahnestock}, P.~{Michel}, D.~C. {Richardson}, S.~P. {Naidu}, D.~J.
  {Scheeres}, A.~F. {Cheng}, A.~S. {Rivkin}, and L.~A.~M. {Benner}.
\newblock {Constraints on the perturbed mutual motion in Didymos due to
  impact-induced deformation of its primary after the DART impact}.
\newblock \emph{MNRAS}, 472\penalty0 (2):\penalty0 1641--1648, Dec. 2017.
\newblock \doi{10.1093/mnras/stx1992}.

\bibitem[{Hyodo} and {Ohtsuki}(2014)]{hyodo2014}
R.~{Hyodo} and K.~{Ohtsuki}.
\newblock {Collisional Disruption of Gravitational Aggregates in the Tidal
  Environment}.
\newblock \emph{ApJ}, 787\penalty0 (1):\penalty0 56, May 2014.
\newblock \doi{10.1088/0004-637X/787/1/56}.

\bibitem[{Hyodo} and {Sugiura}(2022)]{hyodo2022}
R.~{Hyodo} and K.~{Sugiura}.
\newblock {Formation of Moons and Equatorial Ridge around Top-shaped Asteroids
  after Surface Landslide}.
\newblock \emph{ApJL}, 937\penalty0 (2):\penalty0 L36, Oct. 2022.
\newblock \doi{10.3847/2041-8213/ac922d}.

\bibitem[{Hyodo} et~al.(2015){Hyodo}, {Ohtsuki}, and {Takeda}]{hyodo2015}
R.~{Hyodo}, K.~{Ohtsuki}, and T.~{Takeda}.
\newblock {Formation of Multiple-satellite Systems From Low-mass
  Circumplanetary Particle Disks}.
\newblock \emph{ApJ}, 799\penalty0 (1):\penalty0 40, Jan. 2015.
\newblock \doi{10.1088/0004-637X/799/1/40}.

\bibitem[{Irvine}(1965)]{Irvine1965}
W.~M. {Irvine}.
\newblock {Light Scattering by Spherical Particles: Radiation Pressure,
  Asymmetry Factor, and Extinction Cross Section}.
\newblock \emph{Journal of the Optical Society of America}, 55\penalty0
  (1):\penalty0 16, Jan. 1965.

\bibitem[{Jutzi} and {Asphaug}(2015)]{Jutzi2015}
M.~{Jutzi} and E.~{Asphaug}.
\newblock {The shape and structure of cometary nuclei as a result of
  low-velocity accretion}.
\newblock \emph{Science}, 348\penalty0 (6241):\penalty0 1355--1358, June 2015.
\newblock \doi{10.1126/science.aaa4747}.

\bibitem[{Jutzi} et~al.(2008){Jutzi}, {Benz}, and {Michel}]{Jutzi2008a}
M.~{Jutzi}, W.~{Benz}, and P.~{Michel}.
\newblock {Numerical simulations of impacts involving porous bodies. I.
  Implementing sub-resolution porosity in a 3D SPH hydrocode}.
\newblock \emph{Icarus}, 198\penalty0 (1):\penalty0 242--255, Nov. 2008.
\newblock \doi{10.1016/j.icarus.2008.06.013}.

\bibitem[{Jutzi} et~al.(2009){Jutzi}, {Michel}, {Hiraoka}, {Nakamura}, and
  {Benz}]{Jutzi2009b}
M.~{Jutzi}, P.~{Michel}, K.~{Hiraoka}, A.~M. {Nakamura}, and W.~{Benz}.
\newblock {Numerical simulations of impacts involving porous bodies. II.
  Comparison with laboratory experiments}.
\newblock \emph{Icarus}, 201\penalty0 (2):\penalty0 802--813, June 2009.
\newblock \doi{10.1016/j.icarus.2009.01.018}.

\bibitem[{Karjalainen}(2007)]{karjalainen2007}
R.~{Karjalainen}.
\newblock {Aggregate impacts in Saturn's rings}.
\newblock \emph{Icarus}, 189\penalty0 (2):\penalty0 523--537, Aug. 2007.
\newblock \doi{10.1016/j.icarus.2007.02.009}.

\bibitem[{Leinhardt} and {Stewart}(2012)]{Leinhardt2012}
Z.~M. {Leinhardt} and S.~T. {Stewart}.
\newblock {Collisions between Gravity-dominated Bodies. I. Outcome Regimes and
  Scaling Laws}.
\newblock \emph{The Astrophysical Journal}, 745\penalty0 (1):\penalty0 79, Jan.
  2012.
\newblock \doi{10.1088/0004-637X/745/1/79}.

\bibitem[{Leleu} et~al.(2018){Leleu}, {Jutzi}, and {Rubin}]{Leleu2018}
A.~{Leleu}, M.~{Jutzi}, and M.~{Rubin}.
\newblock {The peculiar shapes of Saturn's small inner moons as evidence of
  mergers of similar-sized moonlets}.
\newblock \emph{Nature Astronomy}, 2:\penalty0 555--561, May 2018.
\newblock \doi{10.1038/s41550-018-0471-7}.

\bibitem[{Madeira} and {Giuliatti Winter}(2020)]{Madeira2020}
G.~{Madeira} and S.~M. {Giuliatti Winter}.
\newblock {Effects of immersed moonlets in the ring arc particles of Saturn}.
\newblock \emph{European Physical Journal Special Topics}, 229\penalty0
  (8):\penalty0 1527--1543, May 2020.
\newblock \doi{10.1140/epjst/e2020-900129-5}.

\bibitem[{Madeira} et~al.(2018){Madeira}, {Sfair}, {Mour{\~a}o}, and {Giuliatti
  Winter}]{Madeira2018}
G.~{Madeira}, R.~{Sfair}, D.~C. {Mour{\~a}o}, and S.~M. {Giuliatti Winter}.
\newblock {Production and fate of the G ring arc particles due to Aegaeon
  (Saturn LIII)}.
\newblock \emph{MNRAS}, 475\penalty0 (4):\penalty0 5474--5479, Apr. 2018.
\newblock \doi{10.1093/mnras/sty179}.

\bibitem[{Madeira} et~al.(2022){Madeira}, {Giuliatti Winter}, {Ribeiro}, and
  {Winter}]{Madeira2022a}
G.~{Madeira}, S.~M. {Giuliatti Winter}, T.~{Ribeiro}, and O.~C. {Winter}.
\newblock {Dynamics around non-spherical symmetric bodies - I. The case of a
  spherical body with mass anomaly}.
\newblock \emph{MNRAS}, 510\penalty0 (1):\penalty0 1450--1469, Feb. 2022.
\newblock \doi{10.1093/mnras/stab3552}.

\bibitem[{Madeira} et~al.(2023){Madeira}, {Charnoz}, {Zhang}, {Michel},
  {Hyodo}, {Genda}, and {Giuliatti Winter}]{madeira2022}
G.~{Madeira}, S.~{Charnoz}, Y.~{Zhang}, P.~{Michel}, R.~{Hyodo}, H.~{Genda},
  and S.~{Giuliatti Winter}.
\newblock {Exploring the recycling model of Phobos formation: rubble-pile
  satellites}.
\newblock \emph{in prep.}, 2023.

\bibitem[Margot et~al.(2015)Margot, Pravec, Taylor, Carry, and
  Jacobson]{margot2015}
J.-L. Margot, P.~Pravec, P.~Taylor, B.~Carry, and S.~Jacobson.
\newblock Asteroid systems: binaries, triples, and pairs.
\newblock \emph{Asteroids IV}, 355:\penalty0 374, 2015.

\bibitem[{McMahon} and {Scheeres}(2010)]{McMahon2010}
J.~{McMahon} and D.~{Scheeres}.
\newblock {Detailed prediction for the BYORP effect on binary near-Earth
  Asteroid (66391) 1999 KW4 and implications for the binary population}.
\newblock \emph{Icarus}, 209\penalty0 (2):\penalty0 494--509, Oct. 2010.
\newblock \doi{10.1016/j.icarus.2010.05.016}.

\bibitem[{Meyer-Vernet} and {Sicardy}(1987)]{meyer1987}
N.~{Meyer-Vernet} and B.~{Sicardy}.
\newblock {On the physics of resonant disk-satellite interaction}.
\newblock \emph{Icarus}, 69\penalty0 (1):\penalty0 157--175, Jan. 1987.
\newblock \doi{10.1016/0019-1035(87)90011-X}.

\bibitem[{Michel} et~al.(2016){Michel}, {Cheng}, {K{\"u}ppers}, {Pravec},
  {Blum}, {Delbo}, {Green}, {Rosenblatt}, {Tsiganis}, {Vincent}, {Biele},
  {Ciarletti}, {H{\'e}rique}, {Ulamec}, {Carnelli}, {Galvez}, {Benner},
  {Naidu}, {Barnouin}, {Richardson}, {Rivkin}, {Scheirich}, {Moskovitz},
  {Thirouin}, {Schwartz}, {Campo Bagatin}, and {Yu}]{michel2016}
P.~{Michel}, A.~{Cheng}, M.~{K{\"u}ppers}, P.~{Pravec}, J.~{Blum}, M.~{Delbo},
  S.~F. {Green}, P.~{Rosenblatt}, K.~{Tsiganis}, J.~B. {Vincent}, J.~{Biele},
  V.~{Ciarletti}, A.~{H{\'e}rique}, S.~{Ulamec}, I.~{Carnelli}, A.~{Galvez},
  L.~{Benner}, S.~P. {Naidu}, O.~S. {Barnouin}, D.~C. {Richardson},
  A.~{Rivkin}, P.~{Scheirich}, N.~{Moskovitz}, A.~{Thirouin}, S.~R. {Schwartz},
  A.~{Campo Bagatin}, and Y.~{Yu}.
\newblock {Science case for the Asteroid Impact Mission (AIM): A component of
  the Asteroid Impact \& Deflection Assessment (AIDA) mission}.
\newblock \emph{Advances in Space Research}, 57\penalty0 (12):\penalty0
  2529--2547, June 2016.
\newblock \doi{10.1016/j.asr.2016.03.031}.

\bibitem[{Michel} et~al.(2018){Michel}, {Kueppers}, {Sierks}, {Carnelli},
  {Cheng}, {Mellab}, {Granvik}, {Kestil{\"a}}, {Kohout}, {Muinonen},
  {N{\"a}sil{\"a}}, {Penttila}, {Tikka}, {Tortora}, {Ciarletti}, {H{\'e}rique},
  {Murdoch}, {Asphaug}, {Rivkin}, {Barnouin}, {Bagatin}, {Pravec},
  {Richardson}, {Schwartz}, {Tsiganis}, {Ulamec}, and {Karatekin}]{michel2018}
P.~{Michel}, M.~{Kueppers}, H.~{Sierks}, I.~{Carnelli}, A.~F. {Cheng},
  K.~{Mellab}, M.~{Granvik}, A.~{Kestil{\"a}}, T.~{Kohout}, K.~{Muinonen},
  A.~{N{\"a}sil{\"a}}, A.~{Penttila}, T.~{Tikka}, P.~{Tortora}, V.~{Ciarletti},
  A.~{H{\'e}rique}, N.~{Murdoch}, E.~{Asphaug}, A.~{Rivkin}, O.~{Barnouin},
  A.~C. {Bagatin}, P.~{Pravec}, D.~C. {Richardson}, S.~R. {Schwartz},
  K.~{Tsiganis}, S.~{Ulamec}, and O.~{Karatekin}.
\newblock {European component of the AIDA mission to a binary asteroid:
  Characterization and interpretation of the impact of the DART mission}.
\newblock \emph{Advances in Space Research}, 62\penalty0 (8):\penalty0
  2261--2272, Oct. 2018.
\newblock \doi{10.1016/j.asr.2017.12.020}.

\bibitem[{Michel} et~al.(2020){Michel}, {Ballouz}, {Barnouin}, {Jutzi},
  {Walsh}, {May}, {Manzoni}, {Richardson}, {Schwartz}, {Sugita}, {Watanabe},
  {Miyamoto}, {Hirabayashi}, {Bottke}, {Connolly}, {Yoshikawa}, and
  {Lauretta}]{Michel2020}
P.~{Michel}, R.~L. {Ballouz}, O.~S. {Barnouin}, M.~{Jutzi}, K.~J. {Walsh},
  B.~H. {May}, C.~{Manzoni}, D.~C. {Richardson}, S.~R. {Schwartz}, S.~{Sugita},
  S.~{Watanabe}, H.~{Miyamoto}, M.~{Hirabayashi}, W.~F. {Bottke}, H.~C.
  {Connolly}, M.~{Yoshikawa}, and D.~S. {Lauretta}.
\newblock {Collisional formation of top-shaped asteroids and implications for
  the origins of Ryugu and Bennu}.
\newblock \emph{Nature Communications}, 11:\penalty0 2655, May 2020.
\newblock \doi{10.1038/s41467-020-16433-z}.

\bibitem[{Mignard}(1984)]{Mignard1984}
F.~{Mignard}.
\newblock {Effects of radiation forces on dust particles in planetary rings}.
\newblock In R.~{Greenberg} and A.~{Brahic}, editors, \emph{IAU Colloq. 75:
  Planetary Rings}, pages 333--366, Jan. 1984.

\bibitem[{Naidu} et~al.(2016){Naidu}, {Benner}, {Brozovic}, {Ostro}, {Nolan},
  {Margot}, {Giorgini}, {Magri}, {Pravec}, {Scheirich}, {Scheeres}, and
  {Hirabayashi}]{naidu2016}
S.~{Naidu}, L.~{Benner}, M.~{Brozovic}, S.~J. {Ostro}, M.~C. {Nolan}, J.~L.
  {Margot}, J.~D. {Giorgini}, C.~{Magri}, P.~{Pravec}, P.~{Scheirich}, D.~J.
  {Scheeres}, and M.~{Hirabayashi}.
\newblock {Observations and Characterization of Binary Near-Earth Asteroid
  65803 Didymos, the Target of the AIDA Mission}.
\newblock In \emph{AGU Fall Meeting Abstracts}, pages P52B--02, Dec. 2016.

\bibitem[{Naidu} et~al.(2020){Naidu}, {Benner}, {Brozovic}, {Nolan}, {Ostro},
  {Margot}, {Giorgini}, {Hirabayashi}, {Scheeres}, {Pravec}, {Scheirich},
  {Magri}, and {Jao}]{naidu2020}
S.~P. {Naidu}, L.~A.~M. {Benner}, M.~{Brozovic}, M.~C. {Nolan}, S.~J. {Ostro},
  J.~L. {Margot}, J.~D. {Giorgini}, T.~{Hirabayashi}, D.~J. {Scheeres},
  P.~{Pravec}, P.~{Scheirich}, C.~{Magri}, and J.~S. {Jao}.
\newblock {Radar observations and a physical model of binary near-Earth
  asteroid 65803 Didymos, target of the DART mission}.
\newblock \emph{Icarus}, 348:\penalty0 113777, Sept. 2020.
\newblock \doi{10.1016/j.icarus.2020.113777}.

\bibitem[{Nimmo} and {Matsuyama}(2019)]{nimmo2019}
F.~{Nimmo} and I.~{Matsuyama}.
\newblock {Tidal dissipation in rubble-pile asteroids}.
\newblock \emph{Icarus}, 321:\penalty0 715--721, Mar. 2019.
\newblock \doi{10.1016/j.icarus.2018.12.012}.

\bibitem[{Ostro} et~al.(2006){Ostro}, {Margot}, {Benner}, {Giorgini},
  {Scheeres}, {Fahnestock}, {Broschart}, {Bellerose}, {Nolan}, {Magri},
  {Pravec}, {Scheirich}, {Rose}, {Jurgens}, {De Jong}, and {Suzuki}]{Ostro2006}
S.~J. {Ostro}, J.-L. {Margot}, L.~A.~M. {Benner}, J.~D. {Giorgini}, D.~J.
  {Scheeres}, E.~G. {Fahnestock}, S.~B. {Broschart}, J.~{Bellerose}, M.~C.
  {Nolan}, C.~{Magri}, P.~{Pravec}, P.~{Scheirich}, R.~{Rose}, R.~F. {Jurgens},
  E.~M. {De Jong}, and S.~{Suzuki}.
\newblock {Radar Imaging of Binary Near-Earth Asteroid (66391) 1999 KW4}.
\newblock \emph{Science}, 314\penalty0 (5803):\penalty0 1276--1280, Nov. 2006.
\newblock \doi{10.1126/science.1133622}.

\bibitem[Pajola et~al.(2022)Pajola, Barnouin, Lucchetti, Hirabayashi, Ballouz,
  Asphaug, Ernst, Della~Corte, Farnham, Poggiali, et~al.]{pajola2022}
M.~Pajola, O.~Barnouin, A.~Lucchetti, M.~Hirabayashi, R.-L. Ballouz,
  E.~Asphaug, C.~Ernst, V.~Della~Corte, T.~Farnham, G.~Poggiali, et~al.
\newblock Anticipated geological assessment of the (65803) didymos--dimorphos
  system, target of the dart--liciacube mission.
\newblock \emph{The Planetary Science Journal}, 3\penalty0 (9):\penalty0 210,
  2022.

\bibitem[{Porco} et~al.(2007){Porco}, {Thomas}, {Weiss}, and
  {Richardson}]{porco2007}
C.~C. {Porco}, P.~C. {Thomas}, J.~W. {Weiss}, and D.~C. {Richardson}.
\newblock {Saturn{\textquoteright}s Small Inner Satellites: Clues to Their
  Origins}.
\newblock \emph{Science}, 318\penalty0 (5856):\penalty0 1602, Dec. 2007.
\newblock \doi{10.1126/science.1143977}.

\bibitem[{Pravec} and {Harris}(2007)]{Pravec2007}
P.~{Pravec} and A.~W. {Harris}.
\newblock {Binary asteroid population. 1. Angular momentum content}.
\newblock \emph{Icarus}, 190\penalty0 (1):\penalty0 250--259, Sept. 2007.
\newblock \doi{10.1016/j.icarus.2007.02.023}.

\bibitem[{Pravec} et~al.(2010){Pravec}, {Vokrouhlick{\'y}}, {Polishook},
  {Scheeres}, {Harris}, {Gal{\'a}d}, {Vaduvescu}, {Pozo}, {Barr}, {Longa},
  {Vachier}, {Colas}, {Pray}, {Pollock}, {Reichart}, {Ivarsen}, {Haislip},
  {Lacluyze}, {Ku{\v{s}}nir{\'a}k}, {Henych}, {Marchis}, {Macomber},
  {Jacobson}, {Krugly}, {Sergeev}, and {Leroy}]{Pravec2010}
P.~{Pravec}, D.~{Vokrouhlick{\'y}}, D.~{Polishook}, D.~J. {Scheeres}, A.~W.
  {Harris}, A.~{Gal{\'a}d}, O.~{Vaduvescu}, F.~{Pozo}, A.~{Barr}, P.~{Longa},
  F.~{Vachier}, F.~{Colas}, D.~P. {Pray}, J.~{Pollock}, D.~{Reichart},
  K.~{Ivarsen}, J.~{Haislip}, A.~{Lacluyze}, P.~{Ku{\v{s}}nir{\'a}k},
  T.~{Henych}, F.~{Marchis}, B.~{Macomber}, S.~A. {Jacobson}, Y.~N. {Krugly},
  A.~V. {Sergeev}, and A.~{Leroy}.
\newblock {Formation of asteroid pairs by rotational fission}.
\newblock \emph{Nature}, 466\penalty0 (7310):\penalty0 1085--1088, Aug. 2010.
\newblock \doi{10.1038/nature09315}.

\bibitem[{Pravec} et~al.(2012){Pravec}, {Harris}, {Ku{\v{s}}nir{\'a}k},
  {Gal{\'a}d}, and {Hornoch}]{Pravec2012}
P.~{Pravec}, A.~W. {Harris}, P.~{Ku{\v{s}}nir{\'a}k}, A.~{Gal{\'a}d}, and
  K.~{Hornoch}.
\newblock {Absolute magnitudes of asteroids and a revision of asteroid albedo
  estimates from WISE thermal observations}.
\newblock \emph{Icarus}, 221\penalty0 (1):\penalty0 365--387, Sept. 2012.
\newblock \doi{10.1016/j.icarus.2012.07.026}.

\bibitem[{Richardson} et~al.(2016){Richardson}, {Barnouin}, {Benner}, {Bottke},
  {Campo Bagatin}, {Cheng}, {Hirabayashi}, {Maurel}, {McMahon}, {Michel},
  {Murdoch}, {Naidu}, {Pravec}, {Rivkin}, {Scheeres}, {Scheirich}, {Tsiganis},
  {Zhang}, and {AIDA Dynam. Phys. Prop. Didymos Wkg. Grp.}]{Richardson2016}
D.~C. {Richardson}, O.~S. {Barnouin}, L.~A.~M. {Benner}, W.~F. {Bottke},
  A.~{Campo Bagatin}, A.~F. {Cheng}, M.~{Hirabayashi}, C.~{Maurel}, J.~W.
  {McMahon}, P.~{Michel}, N.~{Murdoch}, S.~P. {Naidu}, P.~{Pravec}, A.~S.
  {Rivkin}, D.~J. {Scheeres}, P.~{Scheirich}, K.~{Tsiganis}, Y.~{Zhang}, and
  {AIDA Dynam. Phys. Prop. Didymos Wkg. Grp.}
\newblock {Dynamical and Physical Properties of 65803 Didymos}.
\newblock In \emph{47th Annual Lunar and Planetary Science Conference}, Lunar
  and Planetary Science Conference, page 1501, Mar. 2016.

\bibitem[{Rubincam}(2000)]{rubincam2000}
D.~P. {Rubincam}.
\newblock {Radiative Spin-up and Spin-down of Small Asteroids}.
\newblock \emph{Icarus}, 148\penalty0 (1):\penalty0 2--11, Nov. 2000.
\newblock \doi{10.1006/icar.2000.6485}.

\bibitem[{Salmon} et~al.(2010){Salmon}, {Charnoz}, {Crida}, and
  {Brahic}]{salmon2010}
J.~{Salmon}, S.~{Charnoz}, A.~{Crida}, and A.~{Brahic}.
\newblock {Long-term and large-scale viscous evolution of dense planetary
  rings}.
\newblock \emph{Icarus}, 209\penalty0 (2):\penalty0 771--785, Oct. 2010.
\newblock \doi{10.1016/j.icarus.2010.05.030}.

\bibitem[{Scheeres}(2007)]{Scheeres2007}
D.~J. {Scheeres}.
\newblock {The dynamical evolution of uniformly rotating asteroids subject to
  YORP}.
\newblock \emph{Icarus}, 188\penalty0 (2):\penalty0 430--450, June 2007.
\newblock \doi{10.1016/j.icarus.2006.12.015}.

\bibitem[{Scheeres}(2015)]{Scheeres2015}
D.~J. {Scheeres}.
\newblock {Landslides and Mass shedding on spinning spheroidal asteroids}.
\newblock \emph{Icarus}, 247:\penalty0 1--17, Feb. 2015.
\newblock \doi{10.1016/j.icarus.2014.09.017}.

\bibitem[{Scheeres} et~al.(2006){Scheeres}, {Fahnestock}, {Ostro}, {Margot},
  {Benner}, {Broschart}, {Bellerose}, {Giorgini}, {Nolan}, {Magri}, {Pravec},
  {Scheirich}, {Rose}, {Jurgens}, {De Jong}, and {Suzuki}]{Scheeres2006}
D.~J. {Scheeres}, E.~G. {Fahnestock}, S.~J. {Ostro}, J.~L. {Margot}, L.~A.~M.
  {Benner}, S.~B. {Broschart}, J.~{Bellerose}, J.~D. {Giorgini}, M.~C. {Nolan},
  C.~{Magri}, P.~{Pravec}, P.~{Scheirich}, R.~{Rose}, R.~F. {Jurgens}, E.~M.
  {De Jong}, and S.~{Suzuki}.
\newblock {Dynamical Configuration of Binary Near-Earth Asteroid (66391) 1999
  KW4}.
\newblock \emph{Science}, 314\penalty0 (5803):\penalty0 1280--1283, Nov. 2006.
\newblock \doi{10.1126/science.1133599}.

\bibitem[{Sugiura} et~al.(2021){Sugiura}, {Kobayashi}, {Watanabe}, {Genda},
  {Hyodo}, and {Inutsuka}]{sugiura2021}
K.~{Sugiura}, H.~{Kobayashi}, S.-i. {Watanabe}, H.~{Genda}, R.~{Hyodo}, and
  S.-i. {Inutsuka}.
\newblock {SPH simulations for shape deformation of rubble-pile asteroids
  through spinup: The challenge for making top-shaped asteroids Ryugu and
  Bennu}.
\newblock \emph{Icarus}, 365:\penalty0 114505, Sept. 2021.
\newblock \doi{10.1016/j.icarus.2021.114505}.

\bibitem[{Terik Daly} et~al.(2022){Terik Daly}, {Ernst}, {Barnouin}, {Gaskell},
  {Palmer}, {Nair}, {Espiritu}, {Hasnain}, {Waller}, {Stickle}, {Nolan},
  {Trigo-Rodr{\'\i}guez}, {Dotto}, {Lucchetti}, {Pajola}, {Ieva}, and
  {Michel}]{Terik2022}
R.~{Terik Daly}, C.~M. {Ernst}, O.~S. {Barnouin}, R.~W. {Gaskell}, E.~E.
  {Palmer}, H.~{Nair}, R.~C. {Espiritu}, S.~{Hasnain}, D.~{Waller}, A.~M.
  {Stickle}, M.~C. {Nolan}, J.~M. {Trigo-Rodr{\'\i}guez}, E.~{Dotto},
  A.~{Lucchetti}, M.~{Pajola}, S.~{Ieva}, and P.~{Michel}.
\newblock {Shape Modeling of Dimorphos for the Double Asteroid Redirection Test
  (DART)}.
\newblock \emph{The Planetary Science Journal}, 3\penalty0 (9):\penalty0 207,
  Sept. 2022.
\newblock \doi{10.3847/PSJ/ac7523}.

\bibitem[{Tiscareno} et~al.(2013){Tiscareno}, {Hedman}, {Burns}, and
  {Castillo-Rogez}]{Tiscareno2013}
M.~S. {Tiscareno}, M.~M. {Hedman}, J.~A. {Burns}, and J.~{Castillo-Rogez}.
\newblock {Compositions and Origins of Outer Planet Systems: Insights from the
  Roche Critical Density}.
\newblock \emph{ApJL}, 765\penalty0 (2):\penalty0 L28, Mar. 2013.
\newblock \doi{10.1088/2041-8205/765/2/L28}.

\bibitem[Tr{\'o}golo et~al.(2021)Tr{\'o}golo, Moreno, Campo~Bagatin, and
  P{\'e}rez~Molina]{trogolo2021}
N.~Tr{\'o}golo, F.~Moreno, A.~Campo~Bagatin, and M.~P{\'e}rez~Molina.
\newblock Analysis of the dynamical evolution of lofted particles around
  (65803) didymos asteroid.
\newblock In \emph{European Planetary Science Congress}, pages EPSC2021--676,
  2021.

\bibitem[{Vokrouhlick{\'y}} and {{\v{C}}apek}(2002)]{Vokrouhlicky2002}
D.~{Vokrouhlick{\'y}} and D.~{{\v{C}}apek}.
\newblock {YORP-Induced Long-Term Evolution of the Spin State of Small
  Asteroids and Meteoroids: Rubincam's Approximation}.
\newblock \emph{Icarus}, 159\penalty0 (2):\penalty0 449--467, Oct. 2002.
\newblock \doi{10.1006/icar.2002.6918}.

\bibitem[Walsh and Jacobson(2015)]{walsh2015}
K.~J. Walsh and S.~A. Jacobson.
\newblock Formation and evolution of binary asteroids.
\newblock \emph{Asteroids IV}, 375, 2015.

\bibitem[{Walsh} et~al.(2008){Walsh}, {Richardson}, and {Michel}]{walsh2008}
K.~J. {Walsh}, D.~C. {Richardson}, and P.~{Michel}.
\newblock {Rotational breakup as the origin of small binary asteroids}.
\newblock \emph{Nature}, 454\penalty0 (7201):\penalty0 188--191, July 2008.
\newblock \doi{10.1038/nature07078}.

\bibitem[{Warner} et~al.(2009){Warner}, {Harris}, and {Pravec}]{Warner2009}
B.~D. {Warner}, A.~W. {Harris}, and P.~{Pravec}.
\newblock {The asteroid lightcurve database}.
\newblock \emph{Icarus}, 202\penalty0 (1):\penalty0 134--146, July 2009.
\newblock \doi{10.1016/j.icarus.2009.02.003}.

\bibitem[{Watanabe} et~al.(2019){Watanabe}, {Hirabayashi}, {Hirata}, {Hirata},
  {Noguchi}, {Shimaki}, {Ikeda}, {Tatsumi}, {Yoshikawa}, {Kikuchi}, {Yabuta},
  {Nakamura}, {Tachibana}, {Ishihara}, {Morota}, {Kitazato}, {Sakatani},
  {Matsumoto}, {Wada}, {Senshu}, {Honda}, {Michikami}, {Takeuchi}, {Kouyama},
  {Honda}, {Kameda}, {Fuse}, {Miyamoto}, {Komatsu}, {Sugita}, {Okada},
  {Namiki}, {Arakawa}, {Ishiguro}, {Abe}, {Gaskell}, {Palmer}, {Barnouin},
  {Michel}, {French}, {McMahon}, {Scheeres}, {Abell}, {Yamamoto}, {Tanaka},
  {Shirai}, {Matsuoka}, {Yamada}, {Yokota}, {Suzuki}, {Yoshioka}, {Cho},
  {Tanaka}, {Nishikawa}, {Sugiyama}, {Kikuchi}, {Hemmi}, {Yamaguchi}, {Ogawa},
  {Ono}, {Mimasu}, {Yoshikawa}, {Takahashi}, {Takei}, {Fujii}, {Hirose},
  {Iwata}, {Hayakawa}, {Hosoda}, {Mori}, {Sawada}, {Shimada}, {Soldini},
  {Yano}, {Tsukizaki}, {Ozaki}, {Iijima}, {Ogawa}, {Fujimoto}, {Ho}, {Moussi},
  {Jaumann}, {Bibring}, {Krause}, {Terui}, {Saiki}, {Nakazawa}, and
  {Tsuda}]{watanabe2019}
S.~{Watanabe}, M.~{Hirabayashi}, N.~{Hirata}, N.~{Hirata}, R.~{Noguchi},
  Y.~{Shimaki}, H.~{Ikeda}, E.~{Tatsumi}, M.~{Yoshikawa}, S.~{Kikuchi},
  H.~{Yabuta}, T.~{Nakamura}, S.~{Tachibana}, Y.~{Ishihara}, T.~{Morota},
  K.~{Kitazato}, N.~{Sakatani}, K.~{Matsumoto}, K.~{Wada}, H.~{Senshu},
  C.~{Honda}, T.~{Michikami}, H.~{Takeuchi}, T.~{Kouyama}, R.~{Honda},
  S.~{Kameda}, T.~{Fuse}, H.~{Miyamoto}, G.~{Komatsu}, S.~{Sugita}, T.~{Okada},
  N.~{Namiki}, M.~{Arakawa}, M.~{Ishiguro}, M.~{Abe}, R.~{Gaskell},
  E.~{Palmer}, O.~S. {Barnouin}, P.~{Michel}, A.~S. {French}, J.~W. {McMahon},
  D.~J. {Scheeres}, P.~A. {Abell}, Y.~{Yamamoto}, S.~{Tanaka}, K.~{Shirai},
  M.~{Matsuoka}, M.~{Yamada}, Y.~{Yokota}, H.~{Suzuki}, K.~{Yoshioka},
  Y.~{Cho}, S.~{Tanaka}, N.~{Nishikawa}, T.~{Sugiyama}, H.~{Kikuchi},
  R.~{Hemmi}, T.~{Yamaguchi}, N.~{Ogawa}, G.~{Ono}, Y.~{Mimasu},
  K.~{Yoshikawa}, T.~{Takahashi}, Y.~{Takei}, A.~{Fujii}, C.~{Hirose},
  T.~{Iwata}, M.~{Hayakawa}, S.~{Hosoda}, O.~{Mori}, H.~{Sawada}, T.~{Shimada},
  S.~{Soldini}, H.~{Yano}, R.~{Tsukizaki}, M.~{Ozaki}, Y.~{Iijima}, K.~{Ogawa},
  M.~{Fujimoto}, T.~M. {Ho}, A.~{Moussi}, R.~{Jaumann}, J.~P. {Bibring},
  C.~{Krause}, F.~{Terui}, T.~{Saiki}, S.~{Nakazawa}, and Y.~{Tsuda}.
\newblock {Hayabusa2 arrives at the carbonaceous asteroid 162173
  Ryugu{\textemdash}A spinning top-shaped rubble pile}.
\newblock \emph{Science}, 364\penalty0 (6437):\penalty0 268--272, Apr. 2019.
\newblock \doi{10.1126/science.aav8032}.

\bibitem[{Yu} et~al.(2018){Yu}, {Michel}, {Hirabayashi}, {Schwartz}, {Zhang},
  {Richardson}, and {Liu}]{Yu2018}
Y.~{Yu}, P.~{Michel}, M.~{Hirabayashi}, S.~R. {Schwartz}, Y.~{Zhang}, D.~C.
  {Richardson}, and X.~{Liu}.
\newblock {The Dynamical Complexity of Surface Mass Shedding from a Top-shaped
  Asteroid Near the Critical Spin Limit}.
\newblock \emph{AJ}, 156\penalty0 (2):\penalty0 59, Aug. 2018.
\newblock \doi{10.3847/1538-3881/aaccf7}.

\bibitem[{Zhang} et~al.(2017){Zhang}, {Richardson}, {Barnouin}, {Maurel},
  {Michel}, {Schwartz}, {Ballouz}, {Benner}, {Naidu}, and {Li}]{Zhang2017}
Y.~{Zhang}, D.~C. {Richardson}, O.~S. {Barnouin}, C.~{Maurel}, P.~{Michel},
  S.~R. {Schwartz}, R.-L. {Ballouz}, L.~A.~M. {Benner}, S.~P. {Naidu}, and
  J.~{Li}.
\newblock {Creep stability of the proposed AIDA mission target 65803 Didymos:
  I. Discrete cohesionless granular physics model}.
\newblock \emph{Icarus}, 294:\penalty0 98--123, Sept. 2017.
\newblock \doi{10.1016/j.icarus.2017.04.027}.

\bibitem[{Zhang} et~al.(2018){Zhang}, {Richardson}, {Barnouin}, {Michel},
  {Schwartz}, and {Ballouz}]{Zhang2018}
Y.~{Zhang}, D.~C. {Richardson}, O.~S. {Barnouin}, P.~{Michel}, S.~R.
  {Schwartz}, and R.-L. {Ballouz}.
\newblock {Rotational Failure of Rubble-pile Bodies: Influences of Shear and
  Cohesive Strengths}.
\newblock \emph{The Astrophysical Journal}, 857\penalty0 (1):\penalty0 15, Apr.
  2018.
\newblock \doi{10.3847/1538-4357/aab5b2}.

\bibitem[{Zhang} et~al.(2021){Zhang}, {Michel}, {Richardson}, {Barnouin},
  {Agrusa}, {Tsiganis}, {Manzoni}, and {May}]{Zhang2021}
Y.~{Zhang}, P.~{Michel}, D.~C. {Richardson}, O.~S. {Barnouin}, H.~F. {Agrusa},
  K.~{Tsiganis}, C.~{Manzoni}, and B.~H. {May}.
\newblock {Creep stability of the DART/Hera mission target 65803 Didymos: II.
  The role of cohesion}.
\newblock \emph{Icarus}, 362:\penalty0 114433, July 2021.
\newblock \doi{10.1016/j.icarus.2021.114433}.

\end{thebibliography}

\end{document}